\documentstyle[amsmath,twoside,amsfonts,epsf]{article}

\def\sech{\mbox{sech\,}}
\def\re{\mbox{Re\,}}
\def\e{{\mathrm{e}}}
\def\im{\mbox{Im\,}}
\def\half{\frac{1}{2}}
\begin{document}

\begin{flushright}{}
{\bf nlin.SI/0004031}
\end{flushright}

\begin{center}

{\LARGE $N $-Soliton Train Interactions and Perturbed\\[10pt] Complex Toda
Chain in Nonlinear Optics.}\\[10pt]

{\large \bf Adiabatic and non-adiabtic aspects.}\\[10pt]

{\bf V. S. Gerdjikov$^a $, I. M. Uzunov$^b $}

\medskip

{\sl $^a $Institute for Nuclear Research and Nuclear Energy,
1784 Sofia, Bulgaria}

\sl $^b ${Institute of Electronics, 1784 Sofia, Bulgaria}

\end{center}

\begin{abstract}
Our previous results on the $N $-soliton interaction in the adiabatic
approximation have been extended. It is shown that the complex Toda chain
(CTC) model is an universal one in the sense that it describes the $N
$-soliton train interactions for all NLEE from the NLS hierarchy. We
derive the perturbed CTC system and show that the small perturbations
affect only the center of mass motion and the global phase of the $N
$-soliton train. A special reduction of CTC describes the interaction of
the sine-Gordon solitons and anti-solitons. The peculiarities of the
interactions in the non-adiabatic cases are outlined.

\end{abstract}


\section{Introduction}

A number of nonlinear optical phenomena in Kerr-like media \cite{Agr} are
described by the perturbed nonlinear Schr\"odinger equation (NLSE) which in
dimensionless variables reads
\begin{equation}\label{eq:nls}
i\frac{\partial u }{ \partial t} + \frac{1 }{ 2 }
\frac{\partial^2 u  }{ \partial x^2} + |u|^2 u(x,t) =iR[u],
\end{equation}
and by its multicomponent generalizations. Below we shall discuss several
perturbations $R[u] $ due to possible linear losses, bandwidth limited
and nonlinear amplifications.

{}For $R[u]=0 $ the NLSE (\ref{eq:nls}) can be solved by means of the
inverse scattering method \cite{ZMNP,FaTa} applied to the
Zakharov-Shabat system $L[u] $. The analytical methods developed allow
one:  i)~to prove that any solution of the NLSE in the limit $t\to\infty
$ tends to a purely solitonic solution; ii)~describe the soliton
interaction in the generic case when all solitons have pair-wise different
velocities.

{}For practical applications one needs to describe the behavior of the
so-called $N $-soliton trains which are solutions of (\ref{eq:nls})
satisfying the initial condition
\begin{equation}\label{eq:inc}
u(x,0) = \sum_{k=1}^{N} u^{\rm 1s}_k(x,t=0).
\end{equation}
Here $u^{\rm 1s}_k(x,t) $ is the one-soliton solution of (\ref{eq:nls}):
\begin{eqnarray}\label{eq:1s}
&&u^{\rm 1s}_k(x,t) = 2\nu _k \e^{i\phi _k}  \sech (z_k(x,t)) ,
\qquad  z_k(x,t)=2\nu _k(x-\xi_k(t)) , \nonumber\\
&&  \phi _k(x,t) =2\mu _k(x-\xi_k(t)) + \delta _k(t),
\qquad \xi_k(t) = 2\mu _kt + \xi_{k,0}, \nonumber\\
&& \delta _k(t) = 2(\mu _k^2 + \nu _k^2) t + \delta _{k,0},
\end{eqnarray}
where $\nu _k $, $\mu _k $, $\xi_k $ and $\delta _k $ are the amplitude,
velocity, position and phase of the $k $-th soliton-like pulse.

Let us first remark that for $N\geq 2 $ the parameters $\mu _k $ and
$\nu_k$ {\em are not} directly related to the discrete spectrum of $L[u]$.
In fact the spectral data of $L[u] $ with $u $ provided by (\ref{eq:inc})
contains not only $2N $ discrete eigenvalues $\lambda _k^\pm =\kappa _k\pm
i\eta_k$, $\eta_k>0 $, $k=1,\dots, N $, but also nonvanishing `radiation'
related to the continuous spectrum of $L[u] $.  However if we take well
separated pulses $|\xi_{k+1,0}-\xi_{k,0}|\simeq r_0\gg 1 $ then the energy
of the `radiation' is of the order of 1\% of the total energy and may well
be neglected. As a result the corresponding $N $-soliton train may be
approximated by an $N $-soliton solution whose interactions in the
generic case (pair-wise different velocities) are well known
\cite{ZMNP,FaTa}.  Even if we approximate $u(x,t) $ by an exact
$N$-soliton solution it is not an easy matter to evaluate the discrete
eigenvalues $\lambda _k^\pm $ and the corresponding `normalization'
constants $C_k^\pm $ of the Jost solutions; this is easy only in the limit
$r_0\to\infty  $, when $\lambda _k^\pm =\mu _k \pm i\nu _k $.

Other difficulties come from the fact that in many of the applications we
need to analyze trains in which: a)~the solitons move with nearly the same
velocities; b)~various perturbations should be taken into account. In such
cases the exact approach based on the inverse scattering method
\cite{ZMNP,FaTa} can not be directly applied and one should look for other
methods \cite{KS,DesChu}.

Our aim is to extend our previous results on the $N $-soliton
interaction in the adiabatic approximation
\cite{KS,DesChu,Arn*94a,Arn*94b,UGL,GKUE,GUED,5a,GEI,GEKDU}.  Firstly we
show that the CTC model is an universal one in the sense that it describes
the $N$-soliton interactions for all NLEE from the NLS hierarchy. We
derive the perturbed CTC (PCTC) system and show that small perturbations
affect only the center of mass motion and the global phase of the
$N$-soliton train.  Special attention is paid to the (anti-) soliton
interaction of the sine-Gordon equation. We show that in the adiabatic
approximation their interaction is described by the Toda chain with
indefinite metric, which is a special reduction of the CTC. We also
outline the peculiarities of the interactions in the non-adiabatic cases.

This paper is an extended version of \cite{VU}.

\section{The $N $-soliton interactions and the Complex Toda Chain}

In \cite{UGL} the quasi-particle approach of Karpman and Solov'ev has
been generalized to any $N>2 $ soliton train in the adiabatic
approximation.  This means that the solitons initially must have nearly
equal amplitudes and velocities and must be well separated, i.e.:
\begin{eqnarray}\label{eq:inc1}
&& |\nu _{k,0} -\nu _{j,0}|\ll \nu _0, \qquad \qquad
|\mu _{k,0} -\mu _{j,0}|\ll \mu _0, \nonumber\\
&& \nu _0(\xi _{k+1,0} -\xi _{k,0}) \simeq r_0\gg 1; \qquad
|\nu _{k,0}-\nu _0|(\xi _{k+1,0} -\xi _{k,0})\ll 1,
\end{eqnarray}
where the additional zeroes in the subscripts in (\ref{eq:inc1}) refer to
the value at $t=0 $ and $\nu _0 $ and $\mu _0 $ are the average
amplitude and velocity of the $N $-soliton train.  The result is a
dynamical system of equations for the $4N $ soliton parameters called the
generalized Karpman-Solov'ev system (GKS).

The GKS is adapted to treat also the perturbed NLS equation. An
exhaustive list of perturbations, relevant for nonlinear optics, which
include linear and nonlinear dispersive and dissipative terms, effects of
sliding filters, amplitude and phase modulations, etc. is studied
\cite{GUED,5a,GEKDU}.  We prove that the linear perturbations affect each
of the solitons separately, while the nonlinear ones lead to additional
interactive terms between neighboring solitons.

Another important step which allowed us to analyze the $N $-soliton
interactions analytically consists in the fact, that under
some additional approximations the GKS reduces to the complex Toda chain
(CTC) with $N $ nodes \cite{GKUE,GUED}:
\begin{equation}\label{eq:ctc}
\frac{d^2 q_k }{ dt^2 } =16\nu _0^2 \left( \e^{q_{k+1}-q_k} -
\e^{q_{k}-q_{k-1}} \right) ,
\end{equation}
where $k=1,\dots, N $ and we assume that $\e^{-q_0}\equiv
\e^{q_{N+1}}\equiv 0$.  The complex dynamical variables $q_k(t) $ are
expressed in terms of the parameters of the $k $-th soliton by:
\begin{eqnarray}\label{eq:q-k}
&& q_{k}(t) = 2i(\mu _0 + i\nu _0) \xi_{k}(t) -i (\delta _k(t) +
\delta (t)) +kQ_0 , \nonumber\\
&&Q_0= \ln 4\nu _0^2 + i\pi, \qquad  \delta (t)= \frac{1 }{N}
\sum_{s=1}^{N} \delta _k(t) .
\end{eqnarray}
The same result has been derived also by using the variational approach
\cite{Arn*94b}; this approach however should be applied with care, see
\cite{Mikh}.

The CTC with $N $ nodes, which may be viewed as a natural generalization
of the corresponding real Toda chain (RTC), provides a very convenient
tool to study the soliton interactions.  Numeric simulations show that CTC
provides an adequate description for the soliton interactions for a wide
class of initial conditions (\ref{eq:inc}) \cite{GUED,5a,GEKDU}.

It is also possible to describe the soliton interactions of other
soliton-type nonlinear equations with different dispersion laws. For any
such equation one can derive the corresponding GKS and CTC model which
could be useful to study the interactions of their $N $-soliton trains,
see \cite{Evst}.

The adiabatic approximation imposes certain restrictions not only on the
soliton parameters (see Eq.~(\ref{eq:inc1})), but also on the spectral
data of the Zakharov-Shabat system $L $. Firstly the discrete eigenvalues
in the upper $\lambda  $-half plane of $L $ must be located in a small
neighborhood around $\lambda _0 $:
\begin{equation}\label{eq:21.1a}
|\lambda _k^+-\lambda _0|^2\simeq \varepsilon ,\qquad
\lambda _0=\sum_{k=1}^{N}\lambda^+_k/N,
\end{equation}
where the small number $\varepsilon  $ determines the overlap between the
neighboring solitons. With the same precision the eigenvalues
$\lambda_k^+$ can be approximated by $2\zeta _k $ where $\zeta _k $ are
the eigenvalues of the Lax matrix for the CTC. Secondly we have a
condition on the initial values of the constants $C_k $ which determine
the initial positions and phases of the pulses. Skipping the details we
get:
\begin{eqnarray}\label{eq:21.1}
&& \ln \left| C_{k+1}^+(0) / C_k^+(0) \right| = 2\nu _0 (\xi_{k+1,0}-
\xi_{k,0}) + {\cal  O}(1) \simeq -2\ln \varepsilon \gg 1.
\end{eqnarray}
The conditions on the discrete eigenvalues $\lambda _k^+ $ allow us also
to explain the universality of the CTC as a model describing the $N
$-soliton interactions. Namely, we claim that the CTC model describes in
the adiabatic approximation the $N $-soliton interactions of all NLEE in
the NLS hierarchy. Indeed, let us consider a higher NLS equation with
dispersion law $F(\lambda ) $, regular in the vicinity of $\lambda _0 $.
Then the time-dependence of $C_k^+ $ is given by
\begin{equation}\label{eq:21.3}
C_k^+(t) = \exp(2iF(\lambda _{k}^+)t) C_k^+(0).
\end{equation}
As a consequence the one-soliton solution will be given by
(\ref{eq:1s}) with $z_k(x,t) $ and $\delta _k(t) $ replaced by
\begin{eqnarray}\label{eq:2.6}
&& z_k(x,t)= 2\nu _k \left(x - \frac{f_{1,k} }{\nu _k}t - \xi_{0,k}
\right) , \\
\label{eq:2.7}
&& \delta _k(t)= \delta _{k,0} + \frac{2 (\mu _kf_{1,k} -\nu _k
f_{0,k})}{ \nu _k }t,
\end{eqnarray}
where $F(\lambda _k^+)= f_{0,k}+if_{1,k} $.
However, due to (\ref{eq:21.1a}) in fact it is enough to take into account
only the first three terms in the Taylor expansion:
\begin{eqnarray}\label{eq:iF}
{}F(\lambda _k^+)&=& F(\lambda _0) + (\lambda _k^+-\lambda _0) \dot{F}_0
+ \half (\lambda _k^+-\lambda _0)^2 \ddot{F}_0 + {\cal O}
(\varepsilon^{3/2}),
\end{eqnarray}
where $\dot{F}_0= (dF/d\lambda )|_{\lambda =\lambda _0} $,
$\ddot{F}_0= (d^2F/d\lambda^2 )|_{\lambda =\lambda _0} $. Comparing
(\ref{eq:21.3}) and (\ref{eq:2.6}) we see that in the adiabatic
approximation only the first three terms in  (\ref{eq:iF}) are important
for the soliton parametrization. This explains why the soliton
interactions for all the equations from the hierarchy is described by the
same universal model: CTC. As an example of higher NLS equations which
also finds important applications in nonlinear optics is the one with
dispersion law $F(\lambda )=2\lambda ^2 + \eta \lambda ^3 $ introduced in
\cite{HasKod}; the GKS for this equation is derived in \cite{Evst}. The
corresponding CTC model is obtained from (\ref{eq:ctc}) by replacing the
coefficient $16\nu _0^2 $ with a factor depending on $F(\lambda_0) $
which can be taken care of by redefining $q_k $. Another possibility is to
choose $F(\lambda )=1/(2\lambda) $ which after additional reduction leads
to the sine-Gordon equation, see Section~4 below.

These results have been further developed by using the fact, that the CTC
is a completely integrable dynamical system with $2N $ degrees of freedom.
The most important consequence of this fact lies in the possibility to
predict the asymptotic behavior of the solitons from the set of their
initial parameters \cite{GKUE,GEI}. Indeed, knowing the initial soliton
parameters we can construct the eigenvalues of the Lax matrix for the CTC
system, which in turn determine the asymptotic velocities of the solitons.

A more detailed study of the solutions of the CTC allowed us to see that
it allows much richer class of asymptotic regimes than the RTC
\cite{GEI,GEKDU}.  We are also able to describe the class of initial
soliton parameters, that lead to each one of these regimes:
i)~asymptotically free propagation of the solitons (the only regime
allowed by RTC); ii)~$N $-soliton bound states with the possibility of a
quasi-equidistant propagation; iii)~mixed asymptotic regimes when part of
the solitons form bound state(s) and the rest separate from them;
iv)~regimes corresponding to the degenerate and singular solutions of the
CTC.

In \cite{GKUE}--\cite{5a}, \cite{GEKDU} a thorough comparison between the
CTC predictions with the numerical solutions of the NLS equation has been
performed and an excellent match has been established for a number of
choices of the initial soliton parameters in each of the regimes listed
above. Special attention has been paid to regime ii) and more specifically
to the possibility for a quasi-equidistant (QED) propagation of all $N $
solitons. A method for the description of the corresponding initial
soliton parameters responsible for this regime has been proposed.

\section{Perturbed NLS and the perturbed CTC}\label{sec:3}

In \cite{GUED} we showed also that the evolution of the $N $-soliton train
(\ref{eq:inc}) of the perturbed NLS equation (\ref{eq:nls}) is  described
by the following dynamical system for the `slow' evolution of the soliton
parameters:
\begin{eqnarray}\label{eq:4.1}
&& \frac{d\nu _k }{ dt } = 16\nu _0^2 (S_{k} - S_{k+1}) + N_k, \\
\label{eq:4.1b}
&& \frac{d\mu _k  }{ dt } = -16\nu _0^2 (C_{k} - C_{k+1}) + M_k, \\
\label{eq:4.1c}
&& \frac{d\xi _k }{ dt } = 2\mu _k + \Xi_k^{(0)} + \Xi_k, \\
\label{eq:4.1d}
&& \frac{d\delta _k }{ dt } = 2(\mu _k^2 +\nu _k^2)  + X_k^{(0)} +
X_k,
\end{eqnarray}
where
\begin{eqnarray}
&& \Xi_k^{(0)} = - 4(S_{k} + S_{k+1}), \\
&& X_k^{(0)} = 2\mu _k\Xi_{k}^{(0)}  + 24\nu _k(C_{k} +C_{k+1}) ,\\
&& C_k(t)-iS_k(t)= -\frac{1}{ 4\nu _0} \e^{q_k(t)-q_{k-1}(t)}.
\end{eqnarray}
The terms $N_k $, \dots , $X_k $ are determined by $R[u] $ below.
As it was shown in \cite{GUED} they contain two types of terms:
a)~`self-interaction' terms depending only on the parameters of the
$k$-th soliton and b)~`nearest-neighbour' interaction terms containing
linear combinations of $S_k $, $C_k $, $S_{k+1} $ and $C_{k+1} $.

In \cite{GUED} we also derived the explicit expressions for $M_k $,
\dots, $X_k $ in terms of the soliton parameters for several
classes of physically important perturbations.
Here we take into account linear and cubic in $u $ perturbations including
the linear and nonlinear gain, third order dispersion (TOD), intrapulse
Raman scattering (IRS) etc, i.e.:
\begin{equation}\label{eq:15p}
R[u] = \sum_{k=0}^{3} c_k {\partial^k u  \over \partial x^k } +
d_0 |u|^2u + {d_1\over 4} u(|u|^2)_x + {d_2\over 4} (|u|^2u_x -u_x^*u^2) ,
\end{equation}
where $c_s $ and $d_s $ are generically complex parameters:
\[
c_s= c_{s0}+ ic_{s1} , \qquad d_s= d_{s0}+ id_{s1} .
\]
Some of these coefficients, namely $c_{01} $, $c_{21} $ and $d_{01} $
can be put to zero without restrictions; this can be done by
conveniently renormalizing $u $, $t $ and $x $.

The next argument which we will use is that the coefficients in
(\ref{eq:15p}) must be small. We start by assuming that they are, like the
terms $S_k $ and $C_k $, of the order of $\varepsilon $; at the same time
the deviations $\tilde{\nu }_k=\nu _k-\nu _0 $, $\tilde{\mu }_k=
\mu_k-\mu_0$  are of the order of $\sqrt{\varepsilon } $. Therefore in
the right hand sides of the equations (\ref{eq:4.1})-(\ref{eq:4.1b}) we
have only terms of the order of $\varepsilon  $, while in the r.h.sides of
(\ref{eq:4.1c})-(\ref{eq:4.1d}) we have also terms of the order of 1 and
$\sqrt{\varepsilon } $. That is why we will simplify the perturbative terms
in the r.h.sides of (\ref{eq:4.1})-(\ref{eq:4.1d}) by taking only the
first few terms in their Taylor expansions, i.e.
\begin{eqnarray}\label{eq:Tayl}
&& Z_k(\nu _k,\mu _k)= Z_{00}+\tilde{\nu }_k Z_{10} +\tilde{\mu }_k
Z_{01} , \qquad  Z_{00} =Z(\nu _0,\mu _0), \nonumber\\
&& Z_{10} = \left. \frac{\partial Z }{\partial \nu _k }
\right|_{\substack{\nu _k=\nu _0\\ \mu _k=\mu _0}}, \qquad
Z_{01} = \left. \frac{\partial Z }{\partial \mu _k }
\right|_{\substack{\nu _k=\nu _0\\ \mu _k=\mu _0}},
\end{eqnarray}
where $Z $ stands for each of the functions $N_k $, $M_k $, $\Xi_k $ and
$X_k $. The explicit expressions for the coefficients in (\ref{eq:Tayl})
for each of the four functions are given in the appendix.

\subsection{Perturbations of order $\varepsilon  $.}\label{ssec:2.1}

Note that due to our assumption about the perturbation constants all
coefficients in $Z_{00} $ are of the order of $\varepsilon  $; so in fact
we have to take into account only $N_{00} $ and $M_{00} $. As a result we
derive the following perturbed version of the CTC model:
\begin{equation}\label{eq:16.3}
\frac{d^2q_k }{ dt^2 } = U_{00} + 16\nu _0^2 \left( \e^{q_{k+1}-q_{k}} -
\e^{q_{k}-q_{k-1}} \right),
\end{equation}
where $U_{00}=-4\mu_0 N_{00}-4\nu _0 (M_{00} +2iN_{00}) $. In deriving
(\ref{eq:16.3}) we took into account also the fact that now $\lambda _0
=\sum_{k=1}^{N} \lambda _k/N $ and consequently $\nu _0 $ become
time-dependent:
\[
{d\lambda _0 \over dt } = {1 \over N }\sum_{k=1}^{N} (M_k + iN_k) \simeq
M_{00} + i N_{00}, \qquad {d\nu _0  \over dt } \simeq N_{00}.
\]

Eq. (\ref{eq:16.3}) can be solved exactly with the result:
\begin{equation}\label{eq:16.4}
q_k(t) = \frac{1 }{2} U_{00}t^2 + V_{00}t+  q^{(0)}_k(t),
\end{equation}
where $q^{(0)}_k(t) $ is a solution of the unperturbed CTC and $V_{00} $
is an arbitrary constant.

If in addition we assume that $V_{00}=0$ and $\mu _0=0 $ then from the
formulae in the appendix we find that:
\begin{equation}\label{eq:17.1}
U_{00}={16\nu _0^2 \over 3 } \left\{ \nu _0 \left[c_{11} + {4  \over 5 }
\nu _0^2 (d_{11} -7c_{31}) \right] -i \left[ 3c_{00} +4\nu _0^2(2 d_{00} -
c_{20})\right] \right\}.
\end{equation}
We remind that $q_k^{(0)} $ is related to the $k $-th soliton parameters
by (\ref{eq:q-k}). Then from (\ref{eq:16.4}) we see that $\re U_{00} $ and
$\re V_{00} $ influence the center of mass motion of the train while $\im
U_{00} $ and $\im V_{00} $ drive the global phase $\delta (t) $. In
particular for the special case when $\re U_{00}=0 $ the effect of
such perturbation will be to make the phase of all solitons oscillate
simultaneously with a rate proportional to $t^2 $.  However the evolution
of the phase and coordinate differences $\delta _{k+1}-\delta _k $ and
$\xi_{k+1} -\xi_k$ will not be influenced.

\subsection{Perturbations of order $\sqrt{\varepsilon } $.}
\label{ssec:2.2}

More complicated and substantially different is the situation when the
perturbative constants become of order $\sqrt{\varepsilon } $. Then the
terms $N_k $, $M_k $ in (\ref{eq:4.1}), (\ref{eq:4.1b}) can be
approximated by linear combinations of $\tilde{\nu }_k $, $\tilde{\mu }_k
$; as a result Equation (\ref{eq:4.1}) acquires the form:
\begin{eqnarray}\label{eq:4.11}
&& {d\nu _k  \over dt } = N_{00} + N_{10}\tilde{\nu }_k +
N_{01}\tilde{\mu }_k  + 16\nu _0^2 (S_{k} - S_{k+1}), \\
\label{eq:4.11b}
&& {d\mu _k  \over dt } = M_{00} + M_{10}\tilde{\nu }_k +
M_{01}\tilde{\mu }_k -16\nu _0^2 (C_{k} - C_{k+1}) , \\
\label{eq:4.11c}
&& {d\xi _k  \over dt } = 2\mu _k + \Xi_{00} + \Xi_{10}\tilde{\nu }_k +
\Xi_{01}\tilde{\mu }_k + \Xi_k^{(0)} , \\
\label{eq:4.11d}
&& {d\delta _k  \over dt } = 2(\mu _k^2 +\nu _k^2) +X_{00} +
X_{10}\tilde{\nu }_k + X_{01}\tilde{\mu }_k + X_k^{(0)},
\end{eqnarray}

Then $N_{00} $ and $M_{00} $ (of order $\sqrt{\varepsilon } $) will be
the leading order terms in (\ref{eq:4.11}), (\ref{eq:4.11b}) while terms
like $c_{0}\tilde{\nu }_k $, $c_{2}\tilde{\mu }_k $ are of the same order
$\varepsilon  $ as the interaction terms (ones with $S_k $ and $C_k $).
The solutions of such PCTC will be qualitatively different from the ones
of CTC and require separate studies.

\section{Sine-Gordon solitons and the CTC}

Here we shortly discuss the (anti-) soliton interactions of the
sine-Gordon equation:
\begin{equation}\label{eq:24.1}
v_{xt} + \sin v(x,t) = \varepsilon R[v] .
\end{equation}
The problem has been attacked long time ago by Spatschek \cite{Spa} and
Karpman and Solov'ev \cite{KS} where the interaction of two (anti-)
solitons has been studied. The one-soliton solution is given by
\begin{equation}\label{eq:v-1s}
v^{\rm 1s} = 2\sigma \arcsin \tanh z \pm \pi, \qquad z = 2\nu (x -
\xi(t)), \nonumber
\end{equation}
where $\xi(t)=t/(4\nu ^2) +\xi_0 $. For $N>2 $ and $R[v]=0 $ in the
adiabatic approximation only the nearest neighbor interactions are
relevant. Then the results of \cite{Spa,KS} generalize to:
\begin{eqnarray}\label{eq:25.5a}
\frac{d\nu _k}{dt} &=&
4\left(\e^{Q_{k +1} - Q_{k}} - \e^{Q_k -Q_{k-1}}   \right), \\
\label{eq:25.5b}
\frac{d\xi _k}{dt} &=& \frac{1}{4\nu _k^2} +
\frac{1}{\nu_0^2} \left(\e^{Q_k -Q_{k-1}} +\e^{Q_{k+1} -Q_{k}}\right),\\
\label{eq:25.5c}
Q_k(t) &=& -2\nu _0 \xi_k(t) + \frac{i\pi}{2} \left(1 - \sigma_k\right),
\end{eqnarray}
where $\sigma _k=1 $ (or $-1$) if at position $k $ we have soliton (or
anti-soliton). Note that the sine-Gordon (anti-) solitons do not have
internal degrees of freedom and are characterized only by their amplitudes
$\nu _k $ and positions $\xi_k $; here we choose $\xi_1 <\xi_2 <\dots
<\xi_N $ and $\nu_0$ is the average amplitude.

We remark also that the sine-Gordon equation is related to the
Zakharov-Shabat system $L[u] $ if we assume $u=v_x/2 $ and request that
$v$ is real. This last condition has important consequences:  i) besides
the (anti-) soliton solutions related to the purely imaginary eigenvalues
$\lambda _k^\pm =\pm i \nu _k $  of $L $, the sine-Gordon equation has
also breather solutions; ii) one can not have two (anti-) solitons moving
with the same speed, i.e.  $\nu _k\neq \nu _j $ for $k\neq j $.

The adiabatic approximations (\ref{eq:21.1a})  mean that $|\nu _k-\nu
_{k+1}|\simeq {\cal  O}(\sqrt{\varepsilon }) $ and in addition we have
(\ref{eq:21.1}). Thus we see that in the right hand side of
(\ref{eq:25.5b}) only the first term is the relevant one; the other two
are of the order of ${\cal  O}(\varepsilon ) $ and can be neglected. If we
now differentiate (\ref{eq:25.5b}) with respect to $t $, use
(\ref{eq:25.5a}) and keep only terms of order $\varepsilon  $ we get:
\begin{equation}\label{eq:26.1}
\frac{d^2Q _k}{dt^2} = \frac{4}{\nu_0 ^2} \left(\e^{Q_{k +1} - Q_{k}}
-  \e^{Q_k -Q_{k-1}}  \right),
\end{equation}
where we replaced $\nu _k $ in the denominator by $\nu_0$.

If all $\sigma _k $ are equal we obtain the real Toda chain. It is also
known from the spectral properties of the Zakharov-Shabat system, that in
the sine-Gordon case we can not have two (anti-) solitons moving with the
same velocities. This means that if we have a sequence of solitons (or
anti-solitons) only then their interaction is purely repulsive and their
asymptotic regime can contain only asymptotically free `particles'. This
facts are compatible with the analytical results on the sine-Gordon
solitons, see \cite{ZMNP,FaTa}.

The model (\ref{eq:26.1}) with generic $\sigma _k $ has been studied by
Kodama and Ye \cite{KoY} and is known as the Toda chain with indefinite
metric. In both cases we can view (\ref{eq:26.1}) as special reduction of
the CTC.  Thus we see that the second involution on the Zakharov-Shabat
needed for the sine-Gordon equation, carries over as a reduction on the
CTC. We should also note that the equations (\ref{eq:25.5c}),
(\ref{eq:26.1}) have solutions with singularities which are periodic in
time, see \cite{KoY,GEI}. The comparizon between the $N $ (anti-) soliton
train dynamics of the sine-Gordon equation and the indefinite metric Toda
chain is yet to be done. This and the studies of the PCTC for the
perturbed sine-Gordon equation will be published elsewhere.

\section{Non-adiabatic Interactions}

If one or more of the `adiabatic conditions' (\ref{eq:inc1}) are
violated then the picture becomes much more complicated. It is possible
that due to strong perturbation some of the soliton pulses come very close
to each other and strongly overlap. Usualy this is combined with strong
deformations of the pulses and substantial emission of `radiation' which
is not accounted for in our model.

To our knowledge there are no effective models which would provide
analytic description of the soliton interactions in such situations.
As main tool giving a physical insight of the soliton dynamics is the
comparison between the numerical solutions to the NLS equation
(\ref{eq:nls}) and the numerical solution of the corresponding
Zakharov-Shabat spectral problem \cite{GUL}.

The method consists in the following: first we solve numerically
the corresponding (perturbed) NLS equation using the standard fast-Fourier
transform (or beam-propagation) method. Then we use the results for the
pulse shape evaluated at a certain distance as an initial potential for
the Zakharov-Shabat eigenvalue problem and determine numerically its
scattering data \cite{BO}. As a result we can determine the time
evolution of the scattering data (including the data, characterizing the
continuous spectrum).

The advantage of this method is the possibility to follow up the
variations of the amplitudes and the velocities of an arbitrary number of
solitons. Here it is possible in a natural way to estimate the energy of
the `radiation', related to the continuous spectrum of $L $.
The disadvantage is in the necessity to know approximately the locations
of the eigenvalues of $L $ at $t=0 $.

Below we use the basic fact that the unperturbed NLS equation is
integrable. As a consequence the evolution of $u(x,t) $ preserves the
spectrum of corresponding Zakharov-Shabat system $L $, which may be
determined from the initial condition $u(x,t=0) $. In particular the
discrete eigenvalues of $L $ will be time-independent since they are
integrals of motion of the NLSE.

If we next consider perturbed NLS equation then generically the
perturbation will violate the integrability. However we assume that
the perturbation is `small' in the sence that it does not destroy
completely the integrability but rather slightly modifies the spectrum of
$L $. In particular the eigenvalues $\lambda _k^\pm $ of $L $ start to
move; here and below the upperscript $+ $ ($- $) means that the
corresponding eigenvalue is such that $\im \lambda _k^+>0 $
($\im \lambda _k^-<0 $).  We remind that the involution $\lambda _k^+=
(\lambda _k^-)^* $ holds, so it is enough to know only the discrete
eigenvalues $\lambda _k^+ $.

To our knowledge there are no explicit criteria which would allow one to
check whether given perturbation is `small' or not.  Some inexplicit
criteria have been formulated in \cite{GIv*92}; in particular they require
that the eigenvalues $\lambda _k^+ $ remain in the upper half-plane (i.e.
$\im \lambda _k^+>0 $ for all $t $), that they do not come close to the
real axis and that they do not coalesce. In terms of the soliton
parameters the second of these condition means that the amplitude of the
pulses should not become very small.

In \cite{GUL} an investigation of the influence of the intrapulse Raman
scattering and the third order dispersion on the discrete eigenvalues of
the Zakharov-Shabat system have been performed by numerical means. Two
qualitatively different initial conditions approximating two-soliton bound
states have been studied.

The first one $u_1(x,t=0)= 2\sech (x) $ correponds to strongly overlapped
soliton pulses; so in this case the adiabatic approximation is not valid.
The spectrum of $L_1 $ (i.e. of $L $ with potential given by $u_1(x,t=0)
$) consists of two eigenvalues in the upper half-plane \cite{SY} with
$\lambda _k^+ = i(k-1/2) $, $k=1,2 $ (and two more in the lower
half-plane). Note also that the distance between these eigenvalues is not
small.

For the second one $u_2(x,t=0)= \sech (x-\delta )+\sech (x+\delta ) $ with
$\delta \simeq 3\div 4 $ the pulses are well separated and the adiabatic
approximation holds. The spectrum of $L_2 $ can not be calculated
presicely; besides the two pairs of eigenvalues it contains also some
small `radiation'.  The eigenvalues $\lambda _k^+ $ can be well
approximated by the eigenvalues of the Lax matrix for the corresponding
CTC. In our case this give:
\begin{equation}\label{eq:la-2}
\lambda _1^+ \simeq {i  \over 2 } \left( 1 + e^{-\delta }\right) \qquad
\lambda _2^+ \simeq {i  \over 2 } \left( 1 - e^{-\delta }\right).
\end{equation}
Note that already for $\delta \simeq 3\div 4 $ the quantity $e^{-\delta }
$ may be considered as small (of the order of $\sqrt{\varepsilon } $;
i.e., these eigenvalues satisfy the adiabaticity condition. Obviously, if
take $\delta  $ to be smaller then the overlap of the solitons grows and
the adiabaticity is violated. As a result (\ref{eq:la-2}) does not give
correct values for the eigenvalues. In the limit of infinitely separated
soliton pulses, i.e. $\delta \to\infty $ the eigenvalues (\ref{eq:la-2})
coalesce. This is related to the fact that the solution to the CTC with
these initial conditions is singular, see \cite{GEI}.

The effect of these two perturbation on the eigenvalues of $L $ are
similar for both types of initial conditions. In what follows we describe
it for the non-adiabatic case with $u_2(x,t=0) $.

{}First we analyse the effect of the third order dispersion (TOD) which is a
Hamiltonian perturbation. The time evolution of the eigenvalues is shown
on Fig.~1 for different strengths $c_{30} $ of TOD. For $c_{30}\leq 0.01
$ it turnes out that the imaginary parts are almost constant while the
real parts are zero.

It is known also that there exist a critical value
$c_{30,\rm cr} =0.022$ \cite{WaiMenLe} where the two-soliton bound state
breaks down. For $c_{30}=0.02 $, which is just below the critical value
two strongly fluctuating real parts show up. Minor fluctuations of the
imaginary parts can also be identified. For $c_{30}=c_{30,\rm cr} $ the
very splitting of the degenerate real parts $\kappa _1=\kappa _2=0 $
appears. After some transition time both real parts attain constant but
different values. The imaginary parts remain almost unchanged. This stage
of deformation of the eigenvalues caused by TOD corresponds to the break
up of the two-soliton bound state into two single, progressively
separating solitons with amplitudes determined by the initial imaginary
parts of the eigenvalues. This behavior of the eigenvalues is consistent
with the previous results.

If $c_{30} $ grows even larger (e.g., $c_{30}\geq 0.03 $), the smaller
imaginary part changes significantly. This second stage of deformation of
the eigenvalues can be described by the ultimate differences between both
real and imaginary parts, respectively, which increase with $c_{30} $. The
change in the imaginary part leads to the creation of `radiation'.

A similar investigation was performed in order to analyze the effect of
dissipative perturbation such as intrapulse Raman scattering on the
eigenvalues. The results are shown in Fig.~2. A remarkable fact to be
mentioned is that unlike for TOD a very weak perturbation ($d_{11}=-0.0004
$) lifts the degeneracy of the real part. The two-soliton bound state
breaks up and two slowly separating solitons with different but constant
amplitudes emerge. The second stage in the deformation (changes in the
imaginary parts) start from $d_{11}=-0.02 $ and differs from the TOD case
in that both imaginary parts change. In contrast to TOD the effect shows
up for considerably weaker perturbations.

The big change of the larger amplitude ($d_{11}< -0.2 $) causes a strong
variation of the corresponding real part due to the amplitude dependence
of the soliton self-frequency shift.

These results clearly illustrate the qualitatively different effect of
Hamiltonian (TOD) and dissipative (IRS) perturbations on the soliton bound
states.

The numeric evaluation of the spectral data of $L $ for each step of
propagation of the soliton train also allows one to control the precision
of the numerical procedure used to solve the NLSE \cite{KMS}.

Another possible effect of the strong perturbations is that the pulses
taken initially to be one-soliton solutions of the NLS, may deform into
the exact travelling-wave solutions of the perturbed NLS equation. Such
effect has been reported in \cite{UMGL} due to the nonlinear gain and
bandwidth limited amplification. Then the perturbed NLS equation goes int
the Ginzburg-Landau (GL) equation whose stationary solutions possess
characteristic phase modulation (chirp).  It is due to this modulation
that the soliton interaction reduces substantially.

\section{Conclusions}

Starting from the GKS model proposed in \cite{Arn*94a,Arn*94b,GKUE,GUED} we
have derived the perturbed CTC system describing the $N $-soliton train
interaction of the perturbed NLS equation in the adiabatic approximation.
{}For small perturbations the PCTC system is again completely integrable and
provides us with an effective tool for analytic study of the asymptotic
regimes of the $N $-soliton trains.  In the non-adiabatic regime we
propose a combined numeric solution of the NLS equation and the
Zakharov-Shabat problem. Finally we mention that these methods can be
applied also to the adiabatic interaction of the multicomponent NLS
equation and its perturbed versions. Such equations describe the
birefringence effects and soliton interactions in multi-mode fibers. We
expect that their soliton interactions will be described by a generalized
CTC-model in which the soliton phases $\delta _k$ are replaced by
`polarization' vectors $\vec{n}_k $, see \cite{VG}.  These results can be
used in soliton-based fiber-optics communications.

It is not difficult to treat also the perturbed sine-Gordon equation and
derive the corresponding perturbed versions of (\ref{eq:26.1}). This and
the study of the interactions of (anti-) solitons with breathers will be
published elsewhere.

{}Finally we stress on the universal character of the CTC in the sense that
it is independent on the dispersion law of the equation whose soliton
interactions it describes.

\section*{Acknowledgements}
We thank Prof. E. Doktorov for useful discussions.
This work was supported in part by contract F-807 with the National
Science Fund of Bulgaria.

\appendix

\section{The coefficients $Z_{\alpha ,\beta } $.}\label{sec:A1}

Here we list the coefficients $Z_{\alpha ,\beta } $ where $Z $ takes the
values $N_k $, $M_k $, $\Xi_k $ and $X_k $ while the pair of indices
$(\alpha ,\beta ) $ is one of the following $(0,0) $, $(1,0) $ and
$(0,1)$, see formula (\ref{eq:Tayl}). With $c_{01}=c_{21}=d_{01}=0 $ (see
the remark after eq. (\ref{eq:15p})) we have:
\begin{eqnarray*}
&& N_{00} = 4\nu _0 \left( \frac{c_{00}}{2} - c_{11}\mu _0 - {2 \over 3}
c_{20} (\nu _0^2 + 3\mu _0^2) + 8 c_{31}\mu _0(\nu _0^2+\mu _0^2)\right.\\
&& \hspace{6cm} \left. +
\frac{4}{3} \nu_0^2 (d_{00}-d_{21}\mu _0) \right), \\
&& N_{10} = 2 \left( c_{00} - 2 c_{11}\mu _0 + (8c_{31}\nu _0\mu _0 -
4c_{20})(\nu _0^2 + \mu _0^2) + 8\nu _0^2 (d_{00}-d_{21}\mu _0)\right), \\
&& N_{01} = - 4\nu _0 \left( c_{11} + 4 c_{20}\mu _0 - 4c_{31}(\nu _0^2 +
3\mu _0^2) - {4 \over 3} d_{21}\nu _0^2 \right),\\
&& M_{00}= - {4 \over 3} \nu _0^2 \left( c_{11} + 4c_{20}\mu _0 -
12c_{31}\left( \mu _0^2 + {7  \over 15 } \nu _0^2\right) + {4 \over 5} \nu
_0^2 d_{11}\right),\\
&& M_{10}= - {8 \over 3 } \nu _0 \left( c_{11} + 4 c_{20} \mu _0 - 12
c_{31}\mu _0^2 - {56 \over 5 } c_{31} \nu _0^2 + {8 \over 5 } d_{11} \nu
_0^2 \right),\\
&& M_{01}= - {4  \over 3 } \nu _0^2 \left( c_{20}- 12 c_{31} \mu _0
\right),\qquad
\Xi_{00} = - c_{10} + 4 c_{30} (\nu _0^2 + 3 \mu _0^2)
- {2  \over 3 } \nu _0^2 d_{10}, \\
&& \Xi_{10} = {4 \over 3} \nu _0 \left( 6 c_{30} - d_{10}\right), \qquad
\Xi_{01} = 24 c_{30}\mu _0, \\
&& X_{00} = c_{01} - 16 c_{30}\mu _0 (\nu _0^2 - \mu _0^2)  +
4\nu _0^2 \left( \mu _0d_{20} - {1  \over 3 } \mu _0d_{10}\right), \\
&& X_{10}= -8\nu _0 \left( 4 c_{30}\mu _0 - d_{20}\mu _0
+ {1  \over 3 } d_{10}\mu _0 \right), \\
&& X_{01}= 48\mu _0^2 c_{30} + 4\nu _0^2 \left( d_{20} - { 1\over 3 }
d_{10}\right).
\end{eqnarray*}

\begin{figure}\label{fig:1}
\caption{Spatial evolution of (a) real and (b) imaginary parts of the
eigenvalues of $2\sech (x) $ for different strengths of TOD:
$c_{30} =0.02 $ (solid), $c_{30} =0.022 $ (dashed), $c_{30} =0.03$
(dotted), $c_{30} =0.05$ (dash-dotted).}
\epsfysize=16cm
\epsfbox{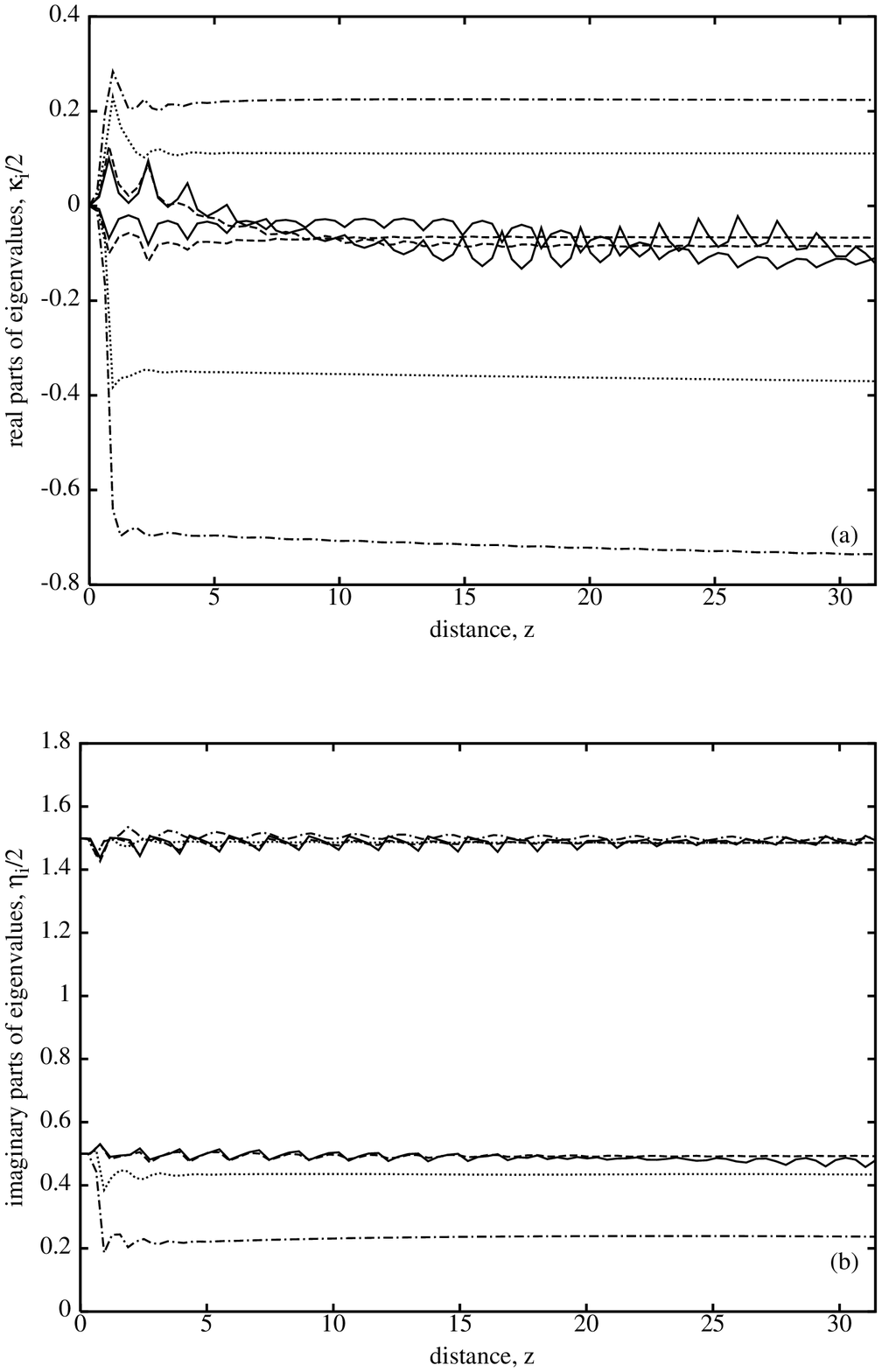}
\vskip -3cm
\end{figure}

\begin{figure}{}\label{fig:2}
\epsfysize=16cm
\caption{Spatial evolution of (a) real and (b) imaginary parts of the
eigenvalues of $2\sech (x) $ for different strengths of IRS:
$d_{11} =-0.0004 $ (smal dashed), $d_{11} =-0.004 $ (solid),
$d_{11} =-0.02$ (small dash-dotted), $d_{11} =-0.04$ (dashed),
$d_{11} =-0.4$ (dashed dotted).}
\epsfbox{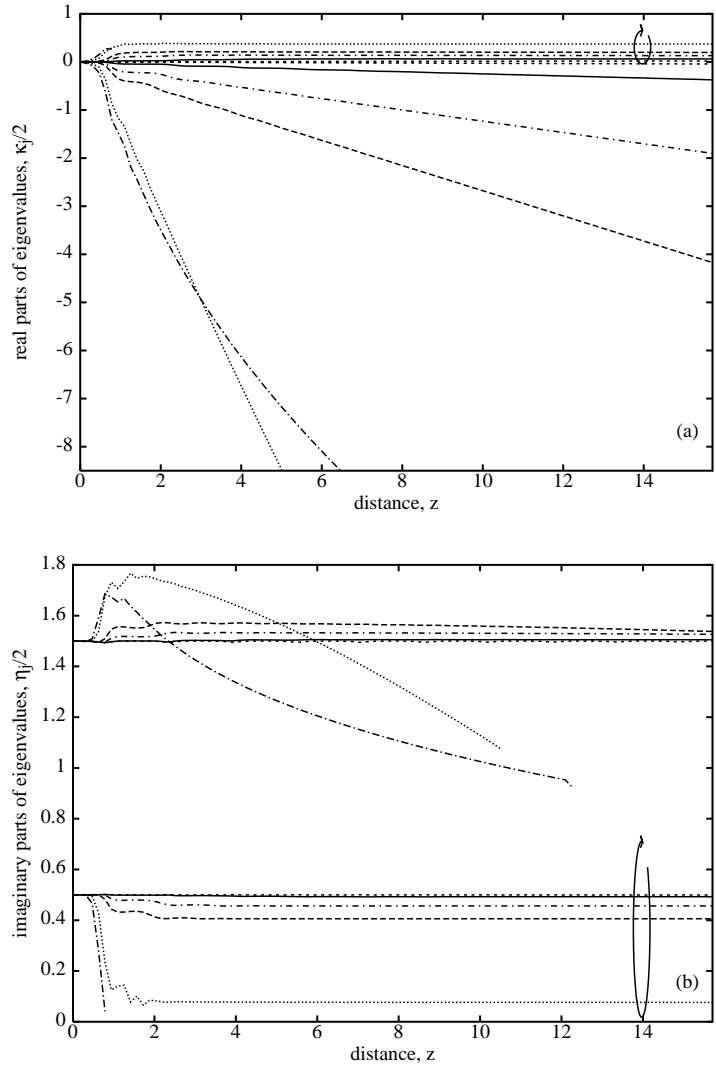}
\end{figure}

\end{document}